\documentclass[prd,twocolumn,superscriptaddress]{revtex4}

\usepackage{caption}
\usepackage{subcaption}
\usepackage{amssymb}
\usepackage{amsfonts}
\usepackage{amsmath}
\usepackage{dcolumn}
\usepackage{bm}
\usepackage[utf8]{inputenc}
\usepackage{graphicx,epsfig}
\usepackage{multirow}
\usepackage{float}
\usepackage[usenames,dvipsnames]{color}
\begin{document}
\title{Bounded Electromagnetic Power Flux at Naked Kasner Singularities}

\author{Franco Fiorini}
\email{franco.fiorini@ib.edu.ar}
\affiliation{Grupo de Comunicaciones Ópticas, Departamento de Ingeniería en Telecomunicaciones, Consejo Nacional de Investigaciones Científicas y Técnicas (CONICET) and Instituto Balseiro (UNCUYO), Centro Atómico Bariloche, Av.~Ezequiel Bustillo 9500, CP8400, San C. de Bariloche, Río Negro, Argentina}
\author{P. A. Gonz\'{a}lez}
\email{pablo.gonzalez@udp.cl} 
\affiliation{Facultad de
Ingenier\'{i}a y Ciencias, Universidad Diego Portales, Avenida Ej\'{e}rcito
Libertador 441, Casilla 298-V, Santiago, Chile.}
\author{Yerko V\'asquez}
\email{yvasquez@userena.cl}
\affiliation{Departamento de F\'isica, Facultad de Ciencias, Universidad de La Serena,\\
Avenida Cisternas 1200, La Serena, Chile.}

\begin{abstract}
In some occasions, it would appear that classical electromagnetic fields could be able to coexist with strong curvature, naked singularities, in the sense that the electromagnetic power flux would be bounded in any open neighborhood of them. We illustrate this rare behavior by considering the naked singularity present in Kasner spacetimes. There, using the Plebanski--Tamm map between Maxwell theory in curved spacetime and the associated description corresponding to an effective anisotropic medium in flat space, we found not only some regular electrostatic field configurations, but also electromagnetic waves whose amplitudes lead to a well defined, bounded power flux as they approach the singularity. Sometimes, in these unusual situations, the singularity resembles a mirror, perfectly reflecting its environs; in other occasions, the existence of a non-null, finite power flux as the waves travel to or from the singular plane, offers a very simple explanation as to how the singularity might absorb or create radiation fields.  
\end{abstract}

\maketitle

\section{INTRODUCTION}

A peculiar coincidence, established first by Tamm \cite{Tamm} and subsequently worked out by Plebanski \cite{Pleb}, allows us to convert the source free Maxwell's equations on an arbitrary curved background, namely
\begin{equation}
F^{\mu\nu}_{\,\,\,\,\,\,\,;\mu}=0\,,\,\,\,\,(\epsilon^{\mu\nu\rho\sigma}F_{\rho\sigma})_{,\mu}=0\,, \label{cov}
\end{equation}
where $F_{\mu\nu}=A_{\nu,\mu}- A_{\mu,\nu}$, and $A_\mu=(\phi,\bar{\mathcal{A}})$ is the 4-potential, into the standard flat-space system 
\begin{eqnarray}
    &&\bar{\nabla} \cdot \bar{\mathcal{D}}=0\,,\,\,\,\,\,\,\,\bar{\nabla} \times \bar{\mathcal{H}}-\partial \bar{\mathcal{D}}/\partial t=0\,,\label{eq:maxwell_mediosfuentescom} \\
   &&\bar{\nabla} \cdot \bar{\mathcal{B}}=0\,,\,\,\,\,\,\,\,\,\,\,\,\,\, \bar{\nabla} \times \bar{\mathcal{E}}+\partial \bar{\mathcal{B}}/\partial t=0\,,
     \label{eq:maxwell_medioscom}
\end{eqnarray}
provided the following constitutive relations hold 
\begin{equation}
\bar{\mathcal{D}}=\textbf{K}\bar{\mathcal{E}}+\bar{\Gamma}\times\bar{\mathcal{H}}\,,\,\,\,\,\,
\bar{\mathcal{B}}=\textbf{K}\bar{\mathcal{H}}-\bar{\Gamma}\times\bar{\mathcal{E}}\,.\label{eq_PT_vector_BDcom}
\end{equation}
The rationale underlying this analogy relies on the fact that both the vector $\bar{\Gamma}$ and the matrix $\textbf{K}$ --the latter playing the role of both relative permittivity and permeability in the constitutive equations-- contain the information about the curved background through the components  \begin{equation}\label{elkyelomega}
 K_{ij}=-\sqrt{-g}\,g^{ij}/g_{00}\,,\,\,\,\,\,\,\,\,
\Gamma_m=g_{0m}/g_{00}\,,
\end{equation}
where latin indexes $i,j,...$ refer to coordinates $x,y,z$ in the 3D flat Euclidean space. Let us take note that this analogy is manifestly non-covariant because it is valid only in Cartesian coordinates
fixed at the material medium. This means that the 
metric components, as well as the inverse metric components appearing in Eq. (\ref{elkyelomega}), are all written in Cartesian
coordinates. In any event, the study of electromagnetic (EM) waves
on a given curved spacetime ends up being equivalent to
the characterization of waves in a 3D material medium
described by the constitutive equations (\ref{eq_PT_vector_BDcom}). This gravito-electromagnetic correspondence is immersed in the realm of the so called analog gravity models \cite{Analog1}, \cite{Tanos}; in the particular context in question, optical phenomena in curved spacetime is emulated through the manipulation of the material properties of the medium encoded in $\textbf{K}$ and $\bar{\Gamma}$, hence bridging the gap between the inherently inaccessible aspects of light associated to the strong regime of the gravitational field and its potential experimental testing in the lab, see \cite{leon}-\cite{RAD2}. 

The gravito-electromagnetic correspondence also offers a fertile ground in order to deal with various conceptual aspects present in virtually any theory of gravity constructed upon metrical notions of space and time, as the existence of spacetime singularities and the potential
formation of Cauchy horizons (creating, thus,
regions with causal violations), see \cite{nos1}-\cite{Barcelo}. In regard to the study of spacetime singularities, the correspondence provides an invaluable, different perspective which we would like to push forward in this article; by its very nature, a curvature singularity is not an event of spacetime, so simple questions as \emph{where} it is located or \emph{when} will occur, are very hard even to formulate \cite{Geroch}. Rather, the existence of curvature singularities in general is revealed through the behavior of causal curves and geodesics in the spacetime. As a matter of fact, the key idea of what now is widely
accepted as the minimum condition for a spacetime to
be defined as singular is that it be timelike (null) geodesically incomplete, i.e., if it contains at least one maximally extended
timelike (null) geodesic whose affine parameter does not
assume values in the full range from $-\infty$ to $\infty$ \cite{hawking_ellis}. On the other hand, Eqs. (\ref{eq:maxwell_mediosfuentescom}), (\ref{eq:maxwell_medioscom}) and  (\ref{eq_PT_vector_BDcom}) invite to see things in a different way, because they show us that the EM field is propagating in a material medium which is covering Euclidean flat space, so the usual unbounded values associated to curvature singularities are endorsed to infinities of the EM field in an otherwise flat, regular 3D space. In this way the spacetime singularities can be viewed as those points or regions \emph{in} Euclidean 3D space where pathologies (as a blow up or indefiniteness) in the EM occur. This is very much alike to what happens in quantum field theory, where divergences (of different sort) take place on top of a flat, regular space. 

Now, it could happen that if the spacetime would have a strong curvature singularity at the point $p$, in the sense that at least one curvature scalar takes unboundedly large values as $x^{\mu}\rightarrow p$, the EM field amplitudes could be bounded and well defined as $x^{\mu}\rightarrow p$ in the flat, Euclidean space underlying Eqs. (\ref{eq:maxwell_mediosfuentescom}), (\ref{eq:maxwell_medioscom}). In this rather unusual situation, we could think about the point $p$ (which from the spacetime perspective has no meaning whatsoever), as if it were \emph{brought back} into the 3D space through a process of completion provided by the (potentially) finite values the EM field adopt there. We actually explored this idea before in the context of a toy model spacetime having a curvature singularity, see \cite{nos1} and \cite{nos2} (or, more synoptically, \cite{nos3}). In the following we will give a number of examples illustrating this picture as comes from the study of light propagating in the background of some metrics solving the vacuum Einstein field equations.

Let us consider the family of static Kasner metrics \cite{Kasner}, \cite{Harvey}
\begin{equation}\label{metkas}
    ds^2=-x^{2p_{1}}dt^2+dx^2+x^{2p_{2}}dy^2+x^{2p_{3}}dz^2\,,
\end{equation}
 which are solutions of the vacuum Einstein field equations if the exponents $p_i$ verify
 \begin{equation}\label{condkas}
p_{1}+ p_{2}+p_{3}=p_{1}^2+p_{2}^2+p_{3}^2=1\,. 
\end{equation}
The exponents can then be written in terms of just one free parameter, namely 
\begin{align}
   & p_{1}=c/(c^2-c+1) \,,\label{parc1}\\
    &p_{2}=(1-c)/(c^2-c+1)\,,\label{parc2}\\
     &p_{3}=c(c-1)/(c^2-c+1)\label{parc3}\,,
\end{align}
with $c$ in the reals. As shown in \cite{Khala}, the metric (\ref{metkas}) can be viewed as the axially-symmetric Weyl metric \cite{Weyl} provided the coordinates $(x,y,z)$ are identified with cylindrical coordinates $(r,\phi,z)$ according to $x\rightarrow r$, $y\rightarrow \phi$. Whether or not one decides to adopt this point of view, the spacetime  (\ref{metkas}), being a vacuum solution of Einstein equations, has identically null scalar curvature $R$ and null squared Ricci invariant $\mathcal{R}^2=R_{\mu \nu}R^{\mu \nu}$. Yet, it gives rise to a Kretschmann scalar $\mathcal{K}=R_{\mu \nu\rho\sigma}R^{\mu \nu\rho\sigma}$ of the form
\begin{equation}\label{kretch}
\mathcal{K}=16\, \frac{p_{2}^2\,p_{3}^2}{p_{2}+p_{3}}\,x^{-4}= \frac{16\,c^2(c-1)^2}{(1+c \,(c-1))^3}\,x^{-4}\,.   \end{equation}
Then, unless $c=0$ or $c=1$ (that is, $p_1=p_3=0$, $p_2=1$ or $p_1=1$, $p_2=p_3=0$, respectively), which is just Minkowski spacetime in disguise, the metric (\ref{metkas}) has a strong curvature singularity at $x=0$, as defined in \cite{Ellis} and \cite{Tipler}. Moreover, the singularity is also naked \cite{Vir1}-\cite{Vir3}. In the $c=0$ case, the Minkowskian form of the interval $ds^2=-dt^2+dx^2+x^2dy^2+dz^2$ is evident if we interpret $x$ as a radial marker in a cylindrical coordinate system $(t,x,y,z)$, provided $y$ acts as an angular coordinate. In turn, when $c=1$, the introduction of Rindler-like coordinates $X=x\cosh(t), T=x\sinh{(t)}$ allows the line element to be converted into the standard Minkowskian form $ds^2=-dT^2+dX^2+dy^2+dz^2$. In the former, $x=0$ is just a coordinate singularity corresponding to the $z$ axis of the cylindrical system; in the latter, $x=0$ corresponds to the Rindler horizon associated to a uniformly accelerated observer in Minkowski space. In both cases, no real curvature singularity exists. Further details on the geodesic structure of the space (\ref{metkas}) can be consulted in \cite{Khala} and \cite{Khala2}.

\section{STATIC FIELDS}

By means of the definitions (\ref{elkyelomega}), the matrix $\textbf{K}$ and the vector $\bar{\Gamma}$ associated to the metric (\ref{metkas}) are
\begin{equation}\label{eq:K}
\textbf{K}=\,x^{-1+2p_{2}+2p_{3}}\,\mbox{diag}(1,x^{-2p_{2}},x^{-2p_{3}} )\,,\,\,\,\,\,\bar{\Gamma} =\bar{0}\,,
\end{equation}
where we have eliminated $p_{1}$ after using (\ref{condkas}). Let us first deal with an electrostatic field on the background (\ref{metkas}). By dropping the time-varying terms in Faraday's and Ampère-Maxwell's laws, taking $\bar{\mathcal{H}}=0$, and combining the constitutive equations (\ref{eq_PT_vector_BDcom}) with $\bar{\nabla} \cdot \bar{\mathcal{D}}=0$, we obtain that electrostatics in the material medium is governed by
\begin{equation}   
    \bar{\nabla} \cdot (\textbf{K}\bar{\nabla}\phi)=0\,,\,\,\,\, 
      (\bar{\nabla} \times\bar{\Gamma})\cdot\bar{\nabla}\phi=0\,, \label{estaticac}
\end{equation}
where $\bar{\mathcal{E}}=-\bar{\nabla}\phi$. According to (\ref{eq:K}) we have that the second equation (\ref{estaticac}) is automatically verified, whilst the first reads
\begin{equation}   
     \partial_{x}(x^{-1+2(p_2+p_3)}\,\partial_{x}\phi )+x^{-1+2p_3}\,\partial^2_{y}\phi+x^{-1+2p_2}\,\partial^2_{z}\phi=0\,. \notag
\end{equation}
For the sake of conciseness, let us take $\phi=\phi(x)$ (this ansatz can be easily extended to the full spatial dependence $\phi(x,y,z)$, along the lines of the analysis performed in \cite{nos1}), which implies 
\begin{eqnarray}
&&\phi(x)=c_{1}\log(x)\,,\,\,\,\,\,\,\,\,\,\,\,\,\,\,\,\,\,\,\,\,\,\,\,\,\,\,\,\,\,\,\,\,\,\,\,\,\,\, \alpha=1\,,\label{pot1}\\
 &&\phi(x)=c_{1}\,x^{2(1-\alpha)}/2(1-\alpha)\,,\,\,\,\,\,\,\,\,\,\, \alpha\neq1\,,\label{pot2}
\end{eqnarray}
where $c_{1}$ is an integration constant and we have defined the parameter $\alpha$ (using (\ref{parc2}) and (\ref{parc3})) according to
\begin{align}
\alpha\stackrel{\text{def}}=p_{2}+p_{3}=(1-c)^2/(c^2-c+1)\,. \label{laalfa}
\end{align}
This parameter ranges in the interval $[0,4/3]$ and vanishes only for
\(c=1\). We see that the above mentioned unconventional coordinatizations of flat space then occur when $\alpha=0$ or $\alpha=1$. When $\alpha=0$, Eq. (\ref{pot2}) applies, and the electrostatic potential is just $\phi(x)=c_{1}x^{2}/2$, corresponding to an electric field $\bar{\mathcal{E}}=- c_{1} \,x\,\hat{x}$, which vanishes at the Rindler horizon. Let us note that curves of constant $x$ in the $T,X$ submanifold defined by constant $y$ and $z$, are described by hyperbolae associated to uniform acceleration in Minkowski space, thus, the electric field is constant for any observer in such a state of motion. Hence, the horizon is behaving as a sort of perfectly conducting surface, because it forces the tangential component of the electric field to be zero there. This is so because in the limit $x\rightarrow 0$, the hyperbolae tend to the straight lines $T=\pm X$, and the electric field becomes purely tangential (and null) on the horizon. In turn, the case $\alpha=1$ is described by (\ref{pot1}), and the potential is very much alike the one corresponding to an infinite (in the $z$ direction) thin wire charged with constant charge density proportional to $c_1$, in accordance with the cylindrical symmetry of the line element in these circumstances. For any other value of  $\alpha$, Eq. (\ref{pot2}) produces an electrostatic field   
\begin{eqnarray}
    &&\bar{\mathcal{E}}=- c_{1} \,x^{1-2\alpha}\,\hat{x}\,,\label{estatico}
\end{eqnarray}
which will be regular at $x=0$ provided $0\leq\alpha\leq 1/2$, being $\alpha=0$ just discussed. Any other value of $\alpha$ within this range will then lead to a well defined, bounded electric field as $x$ approaches the true curvature singularity at $x=0$. As a matter of fact, the energy density $U=\bar{\mathcal{D}}\cdot \bar{\mathcal{E}}=\textbf{K}\,\bar{\mathcal{E}}\cdot\bar{\mathcal{E}}$ scales as $U= c_{1}^2 \,x^{1-2\alpha}$, so the comments just made about the electric field are still in order. We remark that this regular behavior is more the exception than the rule, for any other value of $\alpha$ outside the range  $0\leq\alpha\leq 1/2$ (except for $\alpha=1$), would correspond to curved spacetimes having a singularity at $x=0$ which produces truly divergent electrostatic fields there.

\section{TEM waves}
Let us now move to the study of waves of the form
\begin{eqnarray}
\bar{\mathcal{E}}(\bar{x},t)&=&\bar{E}(\bar{x})\exp\big[i\,k_{0}\,(\bar{k}(\bar{x})\cdot\bar{x}- t)\big]\,,\notag\\
\bar{\mathcal{H}}(\bar{x},t)&=&\bar{H}(\bar{x})\exp\big[i\,k_{0}\,( \bar{k}(\bar{x})\cdot\bar{x}- t)\big]\,,\label{expEyH2}
\end{eqnarray}
where $\bar{k}(\bar{x})$ is the non-dimensional wave vector, $k_{0}$ is the wave number associated to the reference flat vacuum (that is to say, when $\textbf{K}=\textbf{I}$, $\bar{\Gamma}=\bar{0}$), and $\bar{E}(\bar{x}), \bar{H}(\bar{x})$ are the complex, position dependent field amplitudes. Inserting the ansatz (\ref{expEyH2}) into the curl equations (\ref{eq:maxwell_mediosfuentescom})-(\ref{eq:maxwell_medioscom}), and using the constitutive relations (\ref{eq_PT_vector_BDcom}) we get
\begin{align}
&i\,k_{0}^{-1}\,\bar{\nabla}\times\bar{E}=[\bar{\nabla}(\bar{k}\cdot\bar{x})+\bar{\Gamma}]\times\bar{E}-\textbf{K} \bar{H}\,, \label{eq:rot_E_cuasi_planas}\\
&i\,k_{0}^{-1}\,\bar{\nabla}\times\bar{H}=[\bar{\nabla}(\bar{k}\cdot\bar{x})+\bar{\Gamma}]\times\bar{H}+\textbf{K} \bar{E}\label{eq:rot_H_cuasi_planas}\,.
\end{align}
Remembering that, in regard to the spacetime in question, the matrix $\textbf{K}$ and the vector  $\bar{\Gamma}$ are given in Eq. (\ref{eq:K}), we can write down the magnetic field from (\ref{eq:rot_E_cuasi_planas}) according to  
\begin{align}
&\bar{H}=\textbf{K}^{-1}\left[\bar{\nabla}(\bar{k}\cdot\bar{x})\times\bar{E}-i\,k_{0}^{-1}\,\bar{\nabla}\times\bar{E}\right], \label{eq:rot_E_cuasi_planas1}
\end{align}
where the inverse matrix reads
\begin{equation}\label{eq:Kinv}
\textbf{K}^{-1}=\,x^{1-2\alpha}\,\mbox{diag}(1,x^{2p_{2}},x^{2p_{3}})\,.
\end{equation}
With the sole purpose of finding exact solutions allowing us to characterize the nature of the EM fields as they propagate towards $x=0$, let us consider the ansatz $\bar{E}=E(x) \hat{z}$, $\bar{k}=k(x)\hat{x}$. This greatly simplifies the equations because the relevant terms in Eq. (\ref{eq:rot_E_cuasi_planas1}) become
\begin{align}
&\bar{\nabla}\times\bar{E}=(0,-\partial_{x}E,0)\,,\label{calc2}\\
&\bar{\nabla}(\bar{k}\cdot\bar{x})\times\bar{E}=-(0,F\,E,0)\,,
\end{align}
where $F=F(x)=x\partial_{x}k+k$. Therefore, the only non-vanishing component of the magnetic field $\bar{H}=H(x)\hat{y}$ as comes from Eq. (\ref{eq:rot_E_cuasi_planas1}) is
\begin{align}\label{elhy1}
\bar{H}(x)&=-\textbf{K}^{-1} (0,F\, E-i k_{0}^{-1}E^{\prime} ,0)^\top\\
 H(x)&= -x^{1-2p_{3}}\, (F\, E-i k_{0}^{-1}E^{\prime})\,,\label{elhy2}
\end{align}
and, from now on, primes denote differentiation with respect to $x$. Then, under these circumstances, the structure of the EM field is purely transversal. 

The magnetic field thus obtained must be placed in Eq. (\ref{eq:rot_H_cuasi_planas}) with the aim of obtaining a second order, nonlinear differential equation for the electric field $E(x)$. Using (\ref{eq:K}) and (\ref{elhy2}), the ingredients involved in Eq. (\ref{eq:rot_H_cuasi_planas}) are  
\begin{align}
\bar{\nabla}\times \bar{H}&=(0,0,H^{\prime})\label{ing15}\\
 \bar{\nabla}(\bar{k}\cdot\bar{x})\times\bar{H}&= \left(0,0, F\,H\right)\label{ing151}\\
 \textbf{K} \bar{E}&=(0,0,x^{-1+2p_{2}} E)\,, \label{ing152}
\end{align}
thus, Eq. (\ref{eq:rot_H_cuasi_planas}) turns out to be
\begin{equation}
i k_{0}^{-1}H^{\prime}= F H+ x^{-1+2p_{2}}E\,,\label{laecparae}
\end{equation}
where $H^{\prime}$ follows from (\ref{elhy2}):
\begin{equation}\label{elhprima}
H^{\prime}x^{2p_{3}}=(2p_{3}-1)(F E+E^{\prime}/ik_{0})-x[(FE)^{\prime}+E^{\prime \prime}/ik_{0}]\,.
\end{equation}
At this point, we should specify the --otherwise undetermined- wave vector $\bar{k}=k(x)\hat{x}$. Usually, e.g. in the context of a guided wave in flat space, the structure of $\bar{k}$ will ultimately be determined by the introduction of boundary conditions. In the present treatment, we can avoid boundary conditions by paying attention to the geometrical optics (GO) limit of the system (\ref{eq:rot_E_cuasi_planas})-(\ref{eq:rot_H_cuasi_planas}). There, the LHS is negligible, because in that limit, the fields have little spatial variations with respect to $k_0$. Moreover, in the RHS, $\bar{\nabla}(\bar{k}\cdot\bar{x})\approx\bar{k}$, so the light ray trajectories in the GO limit --equivalent to the null geodesics in the spacetime-- can be viewed as coming from the optical Hamiltonian constraint (see \cite{Hamop},\cite{Mackay1})
\begin{equation}\label{hamilton}
\mathrm{H}_{am}\stackrel{\text{def}}=\det(\textbf{K}) -\bar{p}^{\,\intercal} \textbf{K}\,\bar{p}=0\,,\,\,\,\,\,\,\, \bar{p}\stackrel{\text{def}}=\bar{k}+\bar{\Gamma}\,.
\end{equation}
The constraint $\mathrm{H}_{am}=0$ is not other but the dispersion relation, see \cite{nos} for more details. Due to the fact that $\bar{\Gamma}=\bar{0}$ in our case, and bearing in mind Eq. (\ref{eq:K}), $\mathrm{H}_{am}$ becomes  
\begin{equation}\label{hamiltonp}
\mathrm{H}_{am}=\left(x^{-2(1-\alpha)}-k_{1}^2-k_{2}^2\,x^{-2p_{2}}-k_{3}^2\,x^{-2p_{3}}\right)/x^{1-2\alpha}\,.\notag
\end{equation}
The constraint $\mathrm{H}_{am}=0$ demands, by virtue that $\bar{k}=k_1(x)\hat{x}\equiv k(x)\hat{x}$,
\begin{align}\label{elkuno}
k= \pm \sqrt{x^{2(\alpha-1)}}\,=\pm x ^{\alpha-1}\,.
\end{align}
This is the form we are going to adopt for the wave vector in the subsequent analysis. In particular, the function $F=F(x)=x\partial_{x}k+k$ appearing in the differential equation (\ref{laecparae}) will be just
\begin{align}
F(x)=\alpha\, k(x)=\pm\alpha x ^{\alpha-1}\,.\label{laefe}
\end{align}
Notice, because $\alpha \in [0,4/3]$, that the vector $\bar{k}$ and the function $F(x)$ might be both complex when $x<0$.

By means of the expressions (\ref{elhy2}), (\ref{elhprima}), (\ref{laefe}), and after defining
\begin{equation} \label{new}
E(x) = e^{\mp \alpha z} U(x)\,, \,\,\,z=b\, x^{\alpha}/\alpha\,,\,\,\, b=i\, k_{0}\,,
\end{equation}
we can write the Eq. (\ref{laecparae}) to obtain, for all $x\neq0$ 
\begin{equation} \label{equa}
U''(x) + (1-2p_3)x^{-1}U'(x) -b^2 x^{2 \alpha -2} U(x) = 0\,.
\end{equation}
The solution is given by
\begin{equation} \label{eqU}
U(x)= x^{p_3}\left[ a_1 I_{\xi}\left(z\right) +a_2 K_{\xi} \left(z\right) \right]\,, 
\end{equation}
where $a_1$ and $a_2$ are complex integration constants. Here $I_{\xi}$ and $K_{\xi}$ are the modified Bessel functions of the first and second kind, respectively, where $\xi =-p_3/\alpha$, which is generally a non-integer, always different from zero, because we are dealing with curved spacetime. After using the definition
\begin{equation}
K_\xi(z)=\pi
\left[
I_{-\xi}(z)-I_\xi(z)
\right]/2\sin(\pi\xi)\,,
\label{K_identity}
\end{equation}
the function $U(x)$ of Eq. (\ref{eqU}) can be converted into the more symmetric expression 
\begin{equation}\label{eqUsym}
U(x) = x^{p_3}  \left(C_1 I_{\xi}(z)+C_2I_{-\xi}(z)\right)\,,
\end{equation}
where new, redefined integration constants $C_1$ and $C_2$ were introduced.
Once the electric field is known, the characterization of the magnetic field follows at once by means of Eq. (\ref{elhy2}), that is to say
\begin{equation}
\label{elhfinal}
H(x) = x^{1-2p_3} \left( \mp\alpha x ^{\alpha-1} E-E^{\prime}/b \right)\,,
\end{equation}
where we have used Eq. (\ref{laefe}). As a matter of fact, if we plug (\ref{new}) into (\ref{elhfinal}) we get  
\begin{equation}
H(x)=
-x^{1-2p_3}e^{\mp\alpha z}U'(x)/b\,.
\label{elhevaluado}
\end{equation}
Starting from $U(x)$ given in Eq. (\ref{eqUsym}), its derivative $U'(x)$ will come from the identity \cite{AbramowitzStegun}
\begin{equation}
2\,dI_\xi/dz=I_{\xi-1}+I_{\xi+1}\stackrel{\text{def}}=W_\xi\,,\notag
\end{equation}
where we have omitted the argument $z$ in the functions. Then,
\begin{align}
\notag U'(x) x^{1-p_3}
=p_3 \left[C_1I_\xi+C_2I_{-\xi}   
\right]+\frac{bx^{\alpha}}{2}\left[C_1W_{\xi}+C_2
W_{-\xi}\right]\,,
\end{align}
which can be further reduced after using twice the recursive relation $I_{\xi-1}(z)=2\,\xi\, z^{-1}I_{\xi}(z)+I_{\xi+1}(z)$ and the fact that $\xi\alpha=-p_3$:
\begin{equation}\label{laderfunu}
U'(x) x^{1-p_3}=b x^\alpha\left[
C_1I_{\xi+1}+C_2I_{-\xi+1}
\right]+2p_3 C_2 I_{-\xi}\,.
\end{equation}
Substitution of Eq.~(\ref{laderfunu}) into Eq.~(\ref{elhevaluado}) yields
\begin{equation}
H(x)=
-\frac{e^{\mp\alpha z}}{x^{p_3}}\left(x^{\alpha}(C_1I_{\xi+1}+C_2I_{-\xi+1})+\frac{2p_3C_2}{b}I_{-\xi}\right).
\label{elhfinalev}
\end{equation}
The electric and magnetic field components of Eqs. (\ref{new}) and (\ref{elhfinalev}) respectively, permit us to fully characterize the power flux towards or out of the singular plane; the time-average power density in a period $T$ is governed by the Poynting vector 
\begin{align}
    2\langle \bar{S} \rangle &=2T^{-1}\int_0^T \bar{S}(t)\,dt= \operatorname{Re}\left[\bar{\mathcal{E}}\times\bar{\mathcal{H}}^{*} \right]=
     -\hat{x} \operatorname{Re}\left[EH^{*}\right] \label{els1}.\notag
  \end{align}
Next, we will inquire on the potential regularity of the electromagnetic field as well as the power flux when the waves approach the singular plane.

\bigskip

When $x$ is positive, the metric (\ref{metkas}) is real for any choice of Kasner parameters. The behavior of the electric field (\ref{new}) close to the singularity follows from the $z\to0$ expansions involved in the function $U(x)$ of Eq. (\ref{eqUsym}), namely \cite{AbramowitzStegun}
\begin{equation}
I_{\pm\xi}(z)=
\frac{1}{\Gamma(1\pm\xi)}
\left(
\frac{z}{2}
\right)^{\pm\xi}\left[
1+\frac{z^2}{4(1\pm\xi)}+\mathcal O(z^{4})
\right],
\label{eq:I_small_z}
\end{equation}
where $\Gamma$ is the Gamma function. Since $\alpha\xi=-p_3$, one obtains that the electric field (\ref{new}) at the leading order in $x$ goes as
\begin{align}
E(x) &=
E_1(1\mp b\,x^{\alpha}+...)+ E_2 \,x^{2p_3}(1\mp b\,x^{\alpha}+...)\,, \notag \\
E_1 &\stackrel{\text{def}}= C_1 b^{\xi} /(2 \alpha)^{\xi} \Gamma(\xi +1) \,,  \notag \\
E_2 &\stackrel{\text{def}}= C_2 (2 \alpha)^{\xi} / b^{\xi}  \Gamma(1- \xi ) \,.
\label{Uplus_near_zero}
\end{align}
It is clear, thus, that the regularity of the electric field as $x\rightarrow0^+$ depends just on the sign of $p_3$, because $\alpha$ is positive. If we keep both $C_1$ and $C_2$ non null, the electric field will be regular provided $p_3$ is positive, which means $c\in R-(0,1)$, see Eq. (\ref{parc3}). Otherwise, the regularity for all values of $p_3$ will be guaranteed provided $C_2=0$.

The expansion (\ref{eq:I_small_z}) allows us to check that in the limit $x\to0^+$, the magnetic field (\ref{elhfinalev}) at leading order has the structure
\begin{align}\label{hache_near_zero}
H(x)&=H_2(1\mp b\,x^{\alpha})+H_1 x^{2(\alpha-p_3)}+...\,,\\
H_1&\stackrel{\text{def}}= -C_1\, b^{\xi+1}/(2 \alpha)^{\xi+1}\,\Gamma(\xi+2)\,,\notag\\
H_2&\stackrel{\text{def}}= -2 C_2\, p_3\, (2 \alpha)^{\xi}/ b^{\xi+1}\,\Gamma(1-\xi)\,.\notag
\end{align}
Now, the condition in order for the magnetic field to be regular for all $H_1$ and $H_2$ is $\alpha-p_3=p_2>0$, which happens when $c\in(-\infty,1)$, see Eq. (\ref{parc2}). In any other case, the regularity of $H(x)$ would require $H_1=0$ (i.e. $C_1=0$). Hence, we conclude that the regularity of the EM field in the generic case in which both integration constants are non null, requires $c\in(-\infty,0)$.  

The power flux in the vicinity of the singular plane follows easily from (\ref{Uplus_near_zero}) and (\ref{hache_near_zero}). At the leading order we have
\begin{align}\label{flux}
    \langle \bar{S} \rangle_0 = \operatorname{Re}(E_1H_2^{*})\,\hat{x}.
  \end{align}
The power flux is finite as $x\rightarrow 0^+$. For $C_1$ and $C_2$ both different from zero, $\langle \bar{S} \rangle_0$ is non null (provided the phase difference between $E_1$ and $H_2^{*}$ is different from $\pi/2$), because $p_3\neq0$. We see, by writing $C_1 = |C_1|e^{i \theta_1}$ and $C_2 = |C_2| e^{i \theta_2}$, that the power flux can be expressed as
\begin{align}\label{potencero}
\langle \bar{S} \rangle_0 = \hat{x}\,p_3 |C_1| |C_2| \sin (\xi \pi) \sin ( \xi \pi + \Delta \theta)/  \xi \pi k_0 \,,  
\end{align}
where $\Delta \theta = \theta_2 - \theta_1$. We note that $\langle \bar{S} \rangle_0\rightarrow0$ as $k_0\rightarrow\infty$, so no power flux arrives to the singular plane as we go to the GO limit. In other words, the fact that a non null, finite EM power can exists as close of the singular plane as desired, is not a property that can be deduced from the behavior of null geodesics in the spacetime, or equivalently, of light rays on the 3D space. On the other hand, the singularity might act as a perfectly conducting plane as well; this simply requires $C_1=0$, which produces a null electric field and a non-null magnetic field as $x\rightarrow0^+$, which is precisely what happens with the components of a EM wave \emph{at} the surface of a perfect mirror. 

\bigskip

When $x<0$ the metric is generically complex, because the metric coefficients contain non-integer powers of $x^{2p_i}$. In general, we can write 
\begin{align}
x^{2p_i}=|x|^{2p_i}e^{2\pi i p_i}
=i^{4p_i}|x|^{2p_i},
\end{align}
thus the metric for all $x\in R-\{0\}$ reads
\begin{align}\notag
ds^2=-i^{4p_1}|x|^{2p_1}dt^2
+dx^2
+i^{4p_2}|x|^{2p_2}dy^2
+i^{4p_3}|x|^{2p_3}dz^2.
\end{align}
Of course, there exist subclasses of real metrics even for $x<0$; we see that the coefficients will be real provided \(2p_i\in\mathbb Z\), but this condition, in addition to (\ref{parc1})-(\ref{parc3}), just gives us trivial Kasner triplets $p_i$ conducing to flat space.

\section{Final comments}

We have shown that the naked curvature singularity of the static Kasner
spacetime can support EM configurations whose time-averaged
power flux remains finite in an arbitrarily small neighborhood of the
singular surface. The analysis was performed by using the
Plebanski--Tamm formulation, which maps Maxwell theory on the curved
background into an equivalent electromagnetic problem in an anisotropic
medium in flat space. Our exact results make it possible to ascertain whether the
divergence of the Kretschmann curvature scalar at \(x=0\) is necessarily accompanied by divergent EM observables.

Our electrostatic prelude showed that this is not always the case.
For Kasner spacetimes satisfying \(0\leq\alpha\leq1/2\) $(\alpha=0$ relates to Rindler space in which no real singularity exists), the electric field
and its corresponding energy density remain finite as \(x\to0^+\). Thus,
although the spacetimes have a genuine naked curvature singularity, the
associated electrostatic observables do not diverge for this range of Kasner exponents. In turn, for the remaining Kasner parameters compatible with Einstein equations \(1/2<\alpha\leq4/3\) --leaving aside the other non-trivial coordinatization of Minkowski space represented by $\alpha=1$-- the curvature singularity, unsurprisingly, really destroys any notion of electrostatic energy. 

The dynamical sector gives the main result of this work. For TEM modes approaching the singular plane in the region $x>0$, the wave equation admits an exact solution in terms of
modified Bessel functions. It was shown that the time-average power density as $x\rightarrow 0^+$ remains finite, see Eq. (\ref{potencero})
. In some occasions --this is the case after taking $C_1=0$ in Eq. (\ref{eqU})-- the singular plane officiates as a perfect conductor, then, the singularity is not producing any kind of EM fields, rather it merely reflects the already present radiation of its surroundings. More strikingly, other field configurations produce a non null, finite power flux as $x\rightarrow 0^+$, which can be ingoing or outgoing depending on the sign of $\operatorname{Re}(E_1H_2^{*})$,  see Eq. (\ref{flux}). This could actually be interpreted as absorption or emission of radiation by the singularity. 

Finally, it could be understandably argued that the
equations we started from, for instance, Eq. (\ref{equa}) and the very definition of the permittivity/permeability tensor $\textbf{K}$ given in Eq. (\ref{eq:K}), are simply not valid at $x = 0$. This is absolutely true, and we intended to highlight so all along the manuscript. From the 3D perspective, however, $x = 0$ is a region which was left out of the otherwise flat manifold because, normally, divergences of the EM field components occur \emph{there}, that is to say, as $x\rightarrow0^+$. The presence of some bounded EM field configurations in that limit, invites us to think on $x = 0$ more as a sort of inaccessible region \emph{in} 3D Euclidean space for most EM solutions, rather than an excluded-from-physical-reality region. This is very much alike to what happens with the functions $f(x)=1/x$ and $g(x)=\sin{x}/x$ in Euclidean space; while $x=0$ is accessible to $g(x)$ (through a limit point procedure), it is utterly inaccessible to $f(x)$. Of course, for this point of view to properly work, we need bounded EM field configurations as $x\rightarrow0^+$ in the first place. Hence, our findings allow us to conclude that, as far as the propagation of light in singular spaces of this sort is concerned, to rule out the possibility of finite signal emission or absorption --i.e. without total destruction-- by the singularity, is far from being well founded on physical grounds. In that vain, perhaps our work could be helpful to reinforce, from another viewpoint, recent investigations studying the possibility of making sense to trans-singularity, pre-Big Bang physics, see, e.g \cite{singularitycrossing1}-\cite{singularitycrossing6}.

\bigskip

\textbf{\emph{Acknowledgements.}} FF appreciates discussions with Juan Paez on previous, related works, and he is most thankful to Departamento de Física de la Universidad de La Serena, where most of the research leading to this article took place. FF is member of \emph{Carrera del Investigador Cient\'{i}fico} (CONICET), and his work has been supported by CONICET and Instituto Balseiro (UNCUYO). P.A.G. would like to thank the Facultad de Ciencias, Universidad de La Serena for its hospitality. Y. V. acknowledges the financial support of DIDULS/ULS, through the project No PR25538511. \\


\begin{thebibliography}{99}

\bibitem{Tamm} I. E. Tamm, J. Russ. Phys.-Chem. Soc., Phys. Section. \textbf{56} (1924) 248.
\bibitem{Pleb} J. Plebanski, Phys. Rev. \textbf{118} (1960) 1396.
\bibitem{Analog1} C. Barceló, S. Liberati and M. Visser, Living Rev. Rel. \textbf{8} (2005) 12. 
\bibitem{Tanos} D. Faccio, F. Belgiorno, S. Cacciatori, V. Gorini, S. Liberati and U. Moschella, \textit{Analogue Gravity Phenomenology} (2013), Springer.
\bibitem{leon} U. Leonhardt, Science \textbf{312} (2006) 5781.
\bibitem{leon-phil} U. Leonhardt and T. G. Philbin, \textit{Geometry and Light. The Science of Invisibility} (2010), Dover Publications Inc.
\bibitem{leon-phi2} U. Leonhardt and T. G. Philbin,  Prog. Opt. \textbf{53} (2009) 69.
\bibitem{Turner} R. A. Tinguely and A. P. Turner, Comm. Phys. \textbf{3} (2020) 120.
\bibitem{RAD} J. Drori \emph{et al}, Phys. Rev. Lett. \textbf{122} (2019) 010404.
\bibitem{RAD2} L. M. Procopio \emph{et al}, Nature \textbf{655} (2026) 336. 
\bibitem{nos1} F. Fiorini and J. M. Paez, Phys. Rev. \textbf{D111} (2025) 024063. 
\bibitem{nos2} F. Fiorini, S. Hernandez and J. M. Paez, Phys. Lett. \textbf{B865}
\bibitem{Barcelo} C. Barceló, J. Eguia Sánchez, G. García-Moreno and G. Jannes, Eur. Phys. J. \textbf{C82} (2022) 299.
\bibitem{Geroch} R. Geroch, Annals of Phys. \textbf{48} (1968) 526. 
\bibitem{hawking_ellis} S. W. Hawking and G. F. R. Ellis, The large scale structure of spacetime (1973), Cambridge University Press.
\bibitem{nos3} J. M. Paez, F. Fiorini and S. Hernandez,  Astronomische Nachrichten (2026) e70112.
\bibitem{Kasner} E. Kasner, Am. J. of Math. \textbf{43} (1921) 217.
\bibitem{Harvey} A. Harvey, Gen. Rel. Grav. \textbf{21} (1989) 1021; \emph{ibid} \textbf{22} (1990) 1433.  
\bibitem{Khala} I. M. Khalatnikov and S.L. Parnovski, Phys. Lett. \textbf{A66} (1978) 466.
\bibitem{Weyl} H. Weyl, Ann. Phys. \textbf{54} (1917) 117; \emph{ibid} \textbf{59} (1919) 185.
\bibitem{Ellis} G. F. R. Ellis and B. Schmidt, Gen. Rel. Grav. \textbf{8} (1977) 915.
\bibitem{Tipler} F. J. Tipler, Phys. Lett. \textbf{A64} (1977) 8.
\bibitem{Vir1} K.S. Virbhadra, D. Narasimha and S.M. Chitre, Astron. Astrophys. \textbf{337} (1998) 1.
\bibitem{Vir2} K.S. Virbhadra and G.F.R. Ellis, Phys. Rev. \textbf{D65} (2002) 103004.
\bibitem{Vir3} K.S. Virbhadra and C.R. Keeton, Phys. Rev. \textbf{D77} (2008) 124014.
\bibitem{Khala2} S.L. Parnovski, Phys. Lett. \textbf{A73} (1979) 153.
 \bibitem{Hamop} V. Perlick, \textit{Ray optics, Fermat´s Principle, and Applications to General Relativity} (2000), Springer.
\bibitem{Mackay1} T. G. Mackay and A. Lakhtakia, \textit{Electromagnetic Anisotropy and Bianisotropy: A Field Guide} (2019) Sec. ed., World Scientific.
\bibitem{nos} F. Fiorini, S. Hernández and E. Losada, Phys. Rev. \textbf{D104} (2021) 124009. 
\bibitem{AbramowitzStegun} G. E. Andrews, R. Askey and R. Roy, \textit{Special Functions}, in Encyclopedia of Mathematics and its Applications (1999), Cambridge University Press. 
\bibitem{singularitycrossing1} F. D'Ambrosio and C. Rovelli, Class. Quantum Grav. \textbf{35} (2018) 215010.
\bibitem{singularitycrossing2} T. A. Koslowski, F. Mercati and D. Sloan, Phys. Lett. \textbf{B778} (2018) 339. 
\bibitem{singularitycrossing3} F. Mercati, JCAP \textbf{10} (2019) 025.
\bibitem{singularitycrossing4} F. Mercati and D. Sloan, Phys. Rev. \textbf{D106} (2022) 044015. 
\bibitem{singularitycrossing5} M. Adamo and F. Mercati, Phys. Rev. \textbf{D110} (2024) 104033. 
\bibitem{singularitycrossing6} A. Rod Gover, J. Kopiński and A. Waldron, Phys. Rev. Lett. \textbf{133} (2024) 011401.
\end{thebibliography}
\end{document}